\documentclass[twocolumn,prd,nofootinbib,aps,floats,floatfix,amsmath,amssymb,secnumarabic]{revtex4} %

\usepackage{graphicx}
\usepackage[usenames,dvipsnames]{color}
\usepackage{amsmath,amssymb}
\usepackage{slashed}
\usepackage[colorlinks,citecolor=blue]{hyperref}

\newcommand{\be}{\begin{equation}}
\newcommand{\ee}{\end{equation}}
\newcommand{\ba}{\begin{array}}
\newcommand{\ea}{\end{array}}
\newcommand{\bea}{\begin{eqnarray}}
\newcommand{\eea}{\end{eqnarray}}
\newcommand{\sss}{\scriptscriptstyle}
\newcommand{\W}{{\sss W}}
\newcommand{\Z}{{\sss Z}}

\begin{document}
\title{130 GeV dark matter and the Fermi gamma-ray line}
\author{James M.\ Cline}
\email{jcline@physics.mcgill.ca}
\affiliation{Department of Physics, McGill University,
3600 Rue University, Montr\'eal, Qu\'ebec, Canada H3A 2T8}
\begin{abstract}

Based on tentative evidence for a peak in the Fermi gamma-ray spectrum
originating from near the center of the galaxy, it has been suggested
that dark matter of mass $\sim 130$ GeV is annihilating directly into
photons with a cross section  $\sim 24$ times smaller than that needed
for the thermal relic density.  We propose a simple particle physics
model in which the DM is a scalar $X$, with a coupling $\lambda_X
X^2|S|^2$ to a scalar multiplet $S$ carrying electric charge,
which allows  for $XX\to\gamma\gamma$ at one loop due to the virtual
$S$.  We predict  a second monochromatic photon peak at 114 GeV due
to $XX\to\gamma Z$. The $S$ is colored under a hidden sector
SU(N) or QCD to help boost the $XX\to\gamma\gamma$ cross section. 
The analogous coupling $\lambda_h h^2 |S|^2$ to
the Higgs  boson can  naturally increase the partial width for
$h\to\gamma\gamma$ by an amount comparable to its standard model
value, as suggested by recent measurements from CMS.  Due to the
hidden sector SU(N) (or QCD), $S$ binds to its antiparticle to form
$S$-mesons, which will be pair-produced in colliders and  then decay
predominantly to $XX$, $hh$, or to glueballs of the SU(N) which subsequently
decay to photons. The cross section for $X$ on nucleons is close to
the Xenon100 upper limit, suggesting that it
should be discovered soon by direct detection.

\end{abstract}
\maketitle


\begin{figure}[bt]
\centering
\vspace{-0.5cm}
\includegraphics[width=0.45\textwidth]{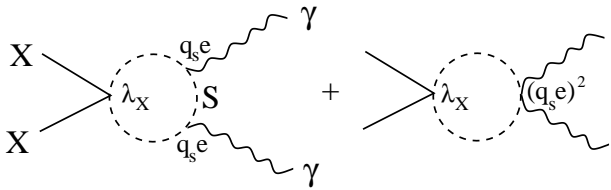}
\caption{Feynman diagrams for the annihilation $XX\to\gamma\gamma$
mediated by virtual $S$.}
\label{ann}
\end{figure}

Refs. \cite{Bringmann:2012vr,Weniger:2012tx} have recently found 
tentative evidence for a 
narrow spectral feature at $E_\gamma =130$ GeV in the Fermi-LAT
\cite{Atwood:2009ez} data (a 4.6$\sigma$ excess, or 3.3$\sigma$ taking into
account the look-elsewhere effect), and have interpreted it as photons from the
annihilation of dark matter (DM) of the same mass.  The Fermi collaboration
does not yet report such a signal, but their most recent upper limit of
$\langle \sigma v\rangle\sim 10^{-27}$cm$^3$s$^{-1}$ (assuming an Einasto
profile) for 130 GeV DM to annihilate
into two photons \cite{Ackermann:2012qk} is consistent with the 
required cross section found in 
\cite{Weniger:2012tx}.  The DM 
interpretation was bolstered in ref.\ \cite{tempel}, which showed that
the two-photon annihilation channel gives a better fit to the feature
than do other final states leading to photons, the others tending to
give a broader peak than is observed.  Ref.\ \cite{Profumo:2012tr}
has suggested that the excess has an astrophysical origin associated
with the Fermi bubble regions, but ref.\
\cite{tempel} claims to locate the spatial regions in
which the signal is maximized, indicating that the strongest
emission is coming from close to the galactic center and not the Fermi
bubble regions.   In this note we adopt the annihilating DM hypothesis
and propose a model which can account for the monochromatic photon
line.\footnote{For an alternative model involving an extra U(1) gauge boson
see \cite{Dudas:2012pb}.  See also \cite{Jackson:2009kg} for an earlier model that can
provide gamma ray lines from DM annihilation.} 

Dark matter (here denoted by $X$) should couple only weakly to photons, if at all,
at tree-level \cite{McDermott:2010pa,Cline:2012is}.  One way to insure
the ``darkness'' of the DM is for it to couple to photons only via
loops.  At one loop, the DM should couple
directly to charged particles $S$.  To make a renormalizable coupling
of this type, both $X$ and $S$ must be bosons, since the stability
of $X$ and the conservation of charge require $X^2$ and $|S|^2$.
This leads us to consider the interactions 
\be
	{\cal L}_{\rm int}  = {\lambda_X\over 2}X^2\, |S|^2 + 
	{\lambda_h}|H|^2\, |S|^2 +
	{\lambda_{hX}\over 2}|H|^2\, X^2
\label{Lint}
\ee
between $X$, the Higgs doublet $H$, and $S$.   The second coupling is
not necessary, but neither is there is any reason to forbid it, and in
fact we will show that it can naturally give rise to an interesting 
enhancement in the $h\to\gamma\gamma$ branching ratio, for the same values 
of
the $S$ mass and charge as needed to explain the Fermi line.  The
third coupling is useful for achieving the correct relic density of 
$X$ \cite{Burgess:2000yq}, as we will discuss.  The stability of
$X$ is insured by the $Z_2$ symmetry $X\to -X$.

{\bf Decays of $S$.} 
It is necessary to make $S$ unstable in order to avoid charged
relics, on whose abundance there are very stringent bounds
from terrestrial searches for anomalous heavy isotopes
\cite{Smith:1982qu,Hemmick:1989ns} and from their effects on big bang nucleosynthesis
\cite{Pospelov:2006sc,Kohri:2006cn}. We will
also find it useful to let $S$ transform under QCD or a hidden SU(N)
gauge symmetry, in order to boost the cross section for
$XX\to\gamma\gamma$.  Suppose $S$ is in the fundamental
representation of SU(N) for definiteness.  If SU(N) is QCD and 
$S$ has charge 4/3, it can decay into right-handed up-type quarks
through the renormalizable operator $\epsilon_{\alpha\beta\gamma}
S_\alpha \bar u_{{\sss R},\beta} u^c_{{\sss R},\gamma}$.
If the SU(N) is exotic, then $S$ could decay into a lighter, neutral
fundamental representation field $T$ and two charged
right-handed fermions through a dimension 5 operator.  For example,
if $S$ has charge $q_S = 2$, the decay into $T+e^+ +e^+$ occurs
via the operator
\be
	{1\over M} T^*_\alpha S^\alpha \bar e^c_{\sss R} e_{\sss R}
\label{decayop}
\ee
which could arise from a renormalizable theory by integrating out
a heavy colored fermion $N_\alpha$ carrying charge 1.  For $q_s=1$,
there is an analogous operator to (\ref{decayop}) involving
left-handed lepton doublets, $T^*_a S^a \bar L_e^c\sigma_2 L_e$, that
mediate the decay $S\to T e^+ \bar\nu$.  We will focus on the 
example (\ref{decayop}) because the larger charge $q_S = 2$ helps
to increase the $XX\to\gamma\gamma$ cross section, and the decays
into electrons can lead to interesting collider signatures.  Of
course, higher-generation leptons can also appear in addition 
to electrons through analogous operators, as well as lepton 
flavor-violating versions, whose contribution to the rare process
$\mu\to e\gamma$ is
suppressed by two loops and two powers of $M$.

Obviously the lifetime of $S$ depends upon the scale $M$ of the
heavy $N_\alpha$  particle.  Estimating the decay rate for
$S\to T ee$ as $m_S^2/(16\pi M^2)$ and demanding the lifetime to 
be less than $10^3$s \cite{Pospelov:2006sc}, we find the limit
$M < 10^{15}$ GeV.  We will show that the relic neutral particle
$T_\alpha$ binds into stable ``baryons'' that make a small
contribution to the total dark matter population.

For simplicity we have assumed that additional couplings of $T$
such as $\lambda_{\sss XT}X^2|T|^2$ and $\lambda_{\sss ST}|S|^2|T|^2$ are small.  The former could
provide a significant annihilation channel $XX\to T T^*$ if 
$m_T< m_X/2$ (the factor of $1/2$ coming from the fact that each
$T$ must hadronize into $T T^*$ bound states) and if $\lambda_{\sss XT}$ is sufficiently large, while the latter has no 
particular impact on the points that follow.

\begin{figure}[tb]
\centering
\vspace{-0.5cm}
\includegraphics[width=0.45\textwidth]{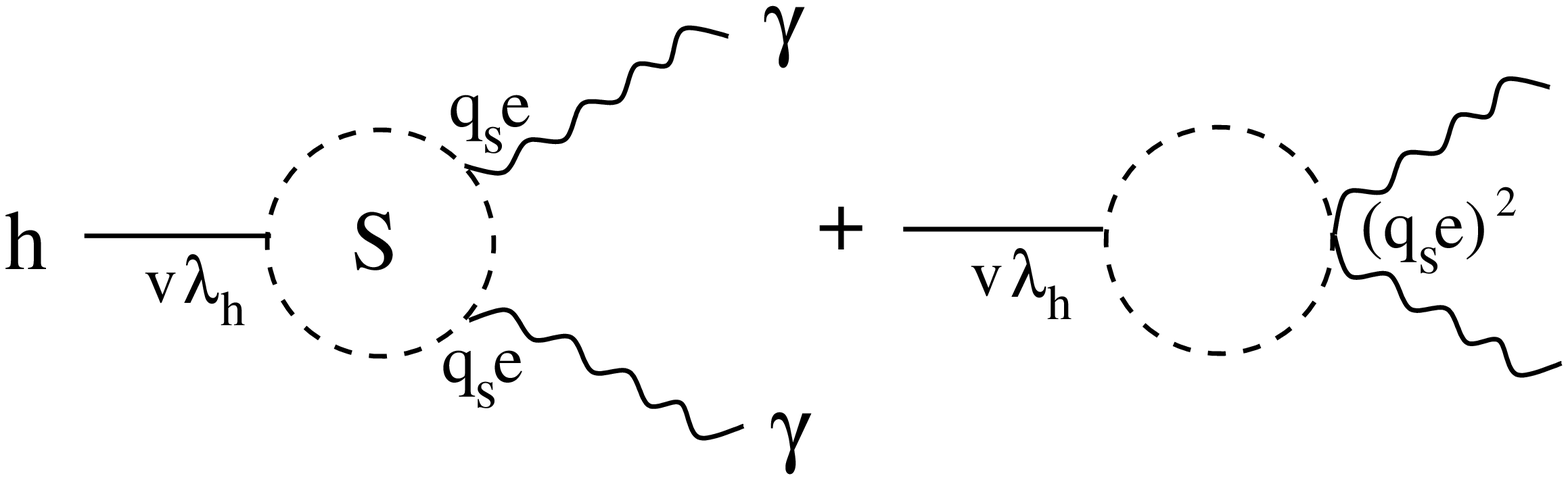}
\caption{Feynman diagrams for the decay  $h\to\gamma\gamma$
mediated by virtual $S$.  $v=246$ GeV is the Higgs VEV.}
\label{decay}
\end{figure}

{\bf Annihilation to two photons.}
The model parameters relevant for the Fermi line are $\lambda_X$,
the mass $m_X$, the charge $q_S$ (in units of $e$), the mass $m_S$,
and the number of colors $N_c$ of QCD or the hidden SU(N) 
gauge group. The
annihilation cross section corresponding to the diagrams of fig.\
\ref{ann} is given by
\be
	\langle\sigma v\rangle = {\sum|{\cal M}|^2\over 64 \pi m_X^2 }
\ee
where the squared matrix element, summed over photon polarizations, is
\be
	\sum|{\cal M}|^2 = {\alpha^2 \over 2\pi^2}\, q_S^4 \lambda_X^2 N_c^2\, 
	\tau^2 A_0^2(\tau)
\label{sumM2}
\ee
with $\tau = m_X^2/m_S^2$ and 
\be
	\tau A_0(\tau) = 1 - \tau^{-1}{\rm arcsin(\sqrt{\tau})^2}
\label{tauAtau}
\ee 
for $\tau \le 1$, which we presume to be the case.   
Eqs.\ (\ref{sumM2},\ref{tauAtau}) can be deduced by comparing to the
well-known result for $h\to\gamma\gamma$ from fig.\ \ref{decay},
in which $\tau \to m_h^2/4 m_S^2$; see for example ref.\ 
\cite{Posch:2010hX}.

\begin{figure}[bt]
\centering
\includegraphics[width=0.4\textwidth]{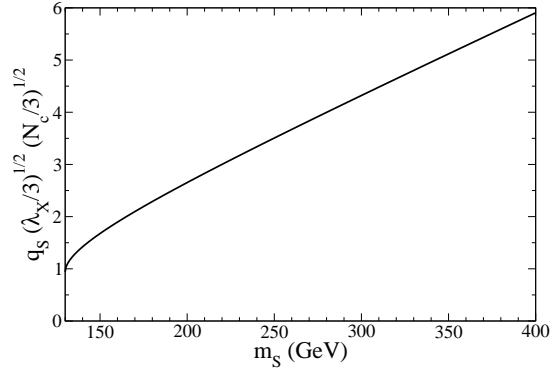}
\caption{Value of $S$ charge $q_S$ (in units of $e$) as a function of
$m_S$, needed to obtain the $X\to\gamma\gamma$ cross section of 
$0.042$ times the standard relic density value
$\langle\sigma v\rangle_0$, assuming $m_X = 130$ GeV.}
\label{sigvrel}
\end{figure}

Ref.\ \cite{Weniger:2012tx} determined that $\langle\sigma v\rangle$ 
should be approximately $0.042$ in units of  the thermal relic 
density value 
$\langle\sigma v\rangle_0 = 1$ pb$\cdot c$, in order to explain the
Fermi gamma-ray line. ({Version 1 of ref.\ \cite{tempel} found 
a larger value,
comparable to $\langle\sigma v\rangle_0$, but this was due to
an error that has now been corrected.})  Taking this value for $\langle\sigma v\rangle$ and 
 $m_X= 130$ GeV, we can find the relation
between $q_S\sqrt{\lambda_X N_c}$ and $m_S$, shown in fig.\
\ref{sigvrel}.   

From fig.\ \ref{sigvrel} we see that even if the coupling
 is rather large, $\lambda_X\sim 3$, and $N_c=3$, 
the $S$ charge is typically greater than 1 (in units of $e$),
and only reaches 1 for $m_S$ close to $m_X$.\footnote{
\label{f2} This limiting
case might be of interest because it allows for the possibility 
of embedding $S$ into an SU(2)$_L$ doublet with the standard
hypercharge assignment for extra Higgs doublets (see arxiv version 1 of this paper for more details).   
However this is not necessary, and
we will focus on the case where $S$ is a singlet of SU(2)$_L$ and 
carries only weak hypercharge.}   At the other extreme of  $q_S \cong 6$, the corresponding
interaction strength $q_S^2\alpha =0.26$ would still be under
perturbative control, indicating that models with $m_S$ up to at least
$400$ GeV are viable.  On the other hand, the rate goes like $q_S^4
\lambda_X^2$, so with $q_S=6$ one could alternatively lower
$\lambda_X$ from $3$ to $0.08$ by taking $m_S\cong m_X$.  Thus it
is not necessary to invoke a very large value of $\lambda_X$.

{\bf Electroweak Precision Constraints.}   $S$ must inherit its
electric charge from weak hypercharge, and therefore it couples
also to the $Z$ boson with strength $q_S e \tan\theta_W$,
where $\theta_W$ is the Weinberg angle.  Such a particle, if 
neutral under SU(2)$_L$, is
unconstrained by precision electroweak constraints since its
contribution to the $\rho$ or $T$ parameters vanishes identically
\cite{Vecchi}, and it contributes only to the $Y$ parameter 
\cite{Arnold:2009ay} which is weakly constrained.

{\bf Annihilation to $Z$ bosons and photons.}  As noted above,
$S$ necessarily
couples to the $Z$ boson as well as to photons. Comparing
the seagull vertices for $|S|^2\gamma\gamma$ and 
$|S|^2\gamma Z$, we can deduce that the cross section for 
$XX\to Z\gamma$ is related to that for $XX\to \gamma\gamma$ by
the factor 
\be
	{\langle\sigma v\rangle_{XX\to \gamma Z}\over
	\langle\sigma v\rangle_{XX\to \gamma\gamma}} =  
	2\tan^2 \theta_W \left(1-{m_Z^2\over4 m_X^2}\right)^{1/2}
	= 0.56
\ee
taking into account the reduced phase space for identical particles
in the case of $XX\to\gamma\gamma$.  Since the former process produces only one photon, its
intensity will be $0.28$ times that of the 2-photon line.  The energy
of this single photon is given by
\be
	E_\gamma = m_X - {m_Z^2\over 4 m_X} = 114{\rm\ GeV}
\ee
We therefore predict that the spectral feature will resolve into
two peaks separated by 16 GeV in energy.  The relative strength of
the peaks could be modified by giving a different SU(2)$_L$
assignment to $S$ so that both components of the doublet become 
electrically charged (see footnote \ref{f2}).  The current Fermi/LAT
limit on $\langle\sigma v\rangle_{XX\to \gamma Z}$ is $2\times
10^{-27}$cm$^3$s$^-1$ \cite{Atwood:2009ez} is compatible with our value.

{\bf Relic density.} 
The cross section for $XX\to\gamma\gamma$ is well below that which is needed
to obtain the right relic density, but this can 
still be achieved
using the $XX\to hh,WW,ZZ$ channels mediated by the interactions
$\frac14\lambda_{hX} h^2 X^2$, $\frac12\lambda_{hX} v h X^2$, 
$\lambda v h^3$ (from the standard model Higgs potential) and
the gauge couplings of the Higgs, with
intermediate Higgs $h$ in the $s$ channel.   
\cite{Burgess:2000yq}.  The cross section for $XX\to hh$ is
\be
	\langle\sigma v\rangle_{hh} = 
	{\lambda_{hX}^2\over 64\pi\, m_X^2}
	\left({r_h\over r_h-2}{\lambda_{hX}\over \lambda} +
{1+r_h/2\over 1-r_h/4}\right)^2
	\sqrt{1 -r_h}
\ee
where $r_h = m_h^2/m_X^2\cong 0.94$  and $\lambda = 0.13$ assuming $m_h = 126$ GeV.  For the
$WW$ final state, we find
\be
	\langle\sigma v\rangle_{\W\W} = {\lambda_{hX}^2\over 8\pi m_X^2}
	 {r_\W^2\over (4-r_h)^2}\left(2 + \left(1-{2\over r_\W}\right)^2\right)
	\sqrt{1 -r_\W}
\ee
where $r_\W = m_W^2/m_X^2$. The cross section 
$\langle\sigma v\rangle_{\Z\Z}$ is the same as 
$\langle\sigma v\rangle_{\W\W}$ with the replacement $r_\W\to r_\Z
\equiv m_Z^2/m_X^2$ 
and a factor of $1/2$ for identical particles in the final state.
Demanding that $\langle\sigma v\rangle_{\rm tot} =
\langle\sigma v\rangle_0$ implies that $|\lambda_{hX}|=0.051$.  
If there are other
significant annihilation channels, this contribution must be
correspondingly reduced so that in general $|\lambda_{hX}|\le 0.05$.
For example if $S$ carries QCD color, then $XX\to$ gluons resulting in 
hadrons can be significant.  There is also the possibility of $XX\to
T T^*$ if the coupling $\lambda_{\sss XT}X^2|T|^2$ is sufficiently 
large and if $m_T < m_X/2$.

\begin{figure}[tb]
\centering
\includegraphics[width=0.45\textwidth]{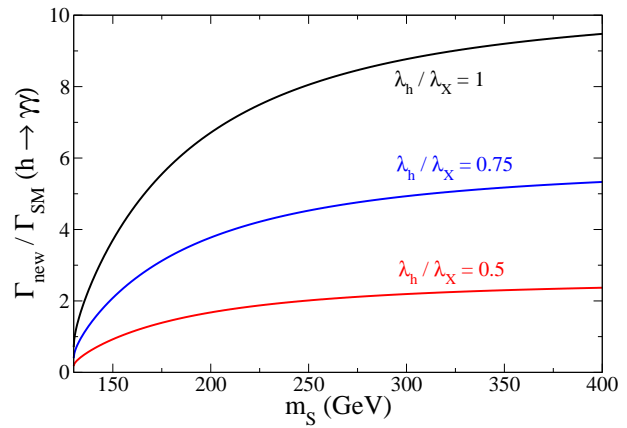}
\caption{Ratio of new and SM contributions to the $h\to \gamma\gamma$
decay width as a function of $m_S$ assuming the relation
in fig.\ \ref{sigvrel}, $m_h=126$ GeV, $m_X=130$ GeV, $\lambda_X =1$, $N_c=3$, 
and $\lambda_h/\lambda_X = 0.5$, $0.75$, $1$ as indicated.
}
\label{h2gg}
\end{figure}

{\bf Direct detection.}  The $\frac12\lambda_{hX} h^2 X^2$ vertex gives
rise to the trilinear interaction $\lambda_{hX} v h X^2$ from 
electroweak symmetry breaking. (Note that the neutral component of the
Higgs doublet is $H^0=\frac{1}{\sqrt{2}}(h+v)$). The Higgs can therefore mediate
scattering of $X$ on nucleons $N$.  The cross section for $XN\to
XN$ elastic scattering is \cite{Barbieri:2006dq}
\be
	\sigma = {f^2\, \lambda_{hX}^2\, m_N^4\over 4\pi\, m_h^4\, m_X^2}
\label{ddsig}
\ee
where $f m_N/v$ is the Higgs-nucleon coupling, with 
$f = \sum_{q=u,d,s} f^p_{Tq} + (2/9)f^p_{TG} = 0.319$ \cite{Giedt:2009mr}.  Using the constraint  
$\lambda_{hX} \lesssim 0.05$ from the preceding relic density determination,
we can evaluate (\ref{ddsig}) to find $\sigma \lesssim 1.3\times
10^{-45}$
cm$^2$.  This is an order of magnitude lower than the 2011
  Xenon100 90\% 
c.l.\ limit of $1.2\times 10^{-44}$ cm$^2$ at $m_X = 130$ GeV
\cite{Aprile:2011hi}, but only 3 times lower than the new limit
of $3\times 10^{-45}$ cm$^2$ that was recently announced
\cite{new_xenon}.
Thus the model could be confirmed or ruled
out in the near future by anticipated improvements \cite{Malling:2011va}
in the direct detection limit.

{\bf Implications for Higgs decays.}
Because of the close similarity between the diagrams of figs.\ 
\ref{ann} and \ref{decay}, there is a simple relation between
$\langle\sigma v\rangle$ and the extra contribution to
$h\to\gamma\gamma$, which is especially interesting in light of
the recent observation by CMS of an upward fluctuation in that
branching ratio, relative to the standard model expectation
\cite{Chatrchyan:2012tx}, assuming of course that the indications of
discovery of the Higgs boson with $m_h\cong 126$ GeV are borne out
\cite{ATLAS:2012ae,Chatrchyan:2012tw,ATLAS:2012ad}.  Specifically,
the squared matrix element for $h\to\gamma\gamma$ is related to 
(\ref{sumM2}) by replacing $\lambda_X\to \lambda_h v$ (where $v$
is the Higgs VEV) and $\tau\to \tilde \tau =  m_h^2/4 m_S^2$.  
The contribution of the charged scalar interferes constructively with
that of the SM if $\lambda_{h} > 0$ \cite{Posch:2010hX}.  
To give an idea its relative size, 
we can express the
extra contribution to the partial width of $h\to\gamma\gamma$ 
(here ignoring interference effects) in
terms of its ratio to the SM contribution (see eq.\ (\ref{tauAtau}),
\be
	{\Gamma_{\rm new}\over \Gamma_{SM}} = 
	q_S^4 \lambda_h^2 N_c^2\, {2\sqrt{2}\, v^2\over A_{SM}^2 G_F\,
m_h^4}\, \tilde\tau^2 A_0^2(\tilde\tau)
\label{grat}
\ee	 
where the SM  amplitude is given by
$A_{SM} = -6.52$ for $m_h=126$ GeV.  For given
values of $\lambda_h/\lambda_X$, $N_c$ and $m_S$
(and assuming that $m_X=130$ GeV)
the ratio (\ref{grat}) is fixed by the Fermi-LAT
cross section, and is $O(1)$ for a wide range of $m_S$ if 
$\lambda_h\sim 0.5\lambda_X$. 
We plot it as a function 
of $m_S$ in fig.\ \ref{h2gg} for several values of
$\lambda_h/\lambda_X$.

{\bf Confinement of $S$ particles.}
Considering the case where $q_s=2$ and 
$S$ decays are mediated by a dimension 5 operator (\ref{decayop}),
generically one would expect the $S$ lifetime to be much larger than
the hadronization time, $1/\Lambda$ for a gauge theory with
confinement mass scale $\Lambda$.  Thus any $S$ particles produced
in a collider will have time to form ``mesonic'' or ``baryonic''
bound states with other $S$ or $T$ particles.  We first discuss the
$S S^*$, $S T^*$ and $T T^*$ mesons, which we denote by $\phi_S$,
$\pi_S$ and $\phi_T$ respectively.   

Through the interaction (\ref{decayop}), $\pi_S$ can decay into two
electrons with a relatively long lifetime.   But $\phi_S$ decays much
faster by its constituents annihilating into $XX$, $hh$,
$\gamma\gamma$, $ZZ$ and $gg$ final states, where  $g$ is the gluon
of the SU(N) theory.  Since there are no lighter colored particles in the
theory, these gluons hadronize into glueballs, which we
denote by $\Omega$. ($\Omega$ is unstable to decay into photons, as we discuss below.)
The partial decay widths can be estimated using $\Gamma \sim
n\langle\sigma v\rangle$ where the density $n$ is the square of the
$S$-meson wave function at the origin, $n\sim (\alpha' m_S/2)^3$, 
$\alpha'$ is the strength of the SU(N) gauge interaction at the scale
$m_S$, and $\sigma$ is the cross section for $S S^*$ scattering into
the desired final state.  In this way we find the partial widths
\be
\Gamma_{\phi_S} \approx {\alpha'^3 m_S\over 8}\left\{\begin{array}{ll}
   \frac{1}{64\pi}\lambda_X^2\sqrt{1-m_X^2/m_S^2}, &
	\phi_S\to XX\\
 & \\
   \frac{1}{64\pi}\lambda_h^2\left(1+{\lambda_h v^2\over
m_h^2-2m_S^2}\right)^2 & \\ 
\qquad\times\sqrt{1-m_h^2/m_S^2},
&	\phi_S\to hh\\
& \\
\alpha^2, & \phi_S\to \gamma\gamma,\, ZZ \\
&\\
\alpha'^2, & \phi_S\to \Omega\,\Omega
\end{array}\right.
\label{widths}
\ee
Because of the large coupling $\lambda_X$, the invisible width for decays 
into $XX$ is typically the most important; even if 
the mass splitting $m_S-m_X$ is small, for example 3 GeV,
$\Gamma_{XX} = 170\, \Gamma_{\gamma\gamma}$ when $\lambda_X=3$.
However, the branching ratio into glueballs $\Omega$ can also
be large and even
dominant, depending  upon the unknown value of $\alpha'$.  The
glueballs can decay into photons via the Euler-Heisenberg interaction
$FFGG/m_S^2$ induced by an $S$ loop.  If the SU(N) confinement scale
$\Lambda$ is below the weak scale, then photons will be the only
kinematically available final states for the glueballs to decay into.
The partial width for $\phi_S\to hh$ can be signficant if $\lambda_h\sim
\lambda_X$.

The $\phi_T$ meson can decay into all the same final states as the
$\phi_S$  by first going into two gluons of the SU(N) theory, which
turn into two other particles by going through a loop of $S$.  These
amplitudes thus occur at two loops.    They can be dominated by a
one-loop contribution due to the possible interaction
$\lambda|S|^2|T|^2$.  Thus we expect the $\phi_T$ partial widths to
be proportional to those of $\phi_S$, with an extra loop and coupling
constant suppression.

The $S$ and $T$ particles can also bind together into baryons of the
SU(N).  For example if $N=3$, we have the $SSS$, $SST$, $STT$ and $TTT$
baryons, which can decay to states with fewer $S$'s via $S\to T+2e$ until
reaching the stable, neutral $TTT$ state.  The former would be stable
charged relics in the absence of this decay mechanism.   The latter is a
dark matter candidate, but its annihilation cross section is too large for
it to provide a significant contribution to the total DM density.  The 
$TTT$ annihilation cross  section is of order $\alpha'^2/(3m_T)^2$, which
for $\alpha' = 0.1$ (similar to the strength of QCD at these energy
scales) is some 2 orders of magnitude larger than needed to get the
correct thermal relic density.

\begin{figure}[tb]
\centering
\includegraphics[width=0.4\textwidth]{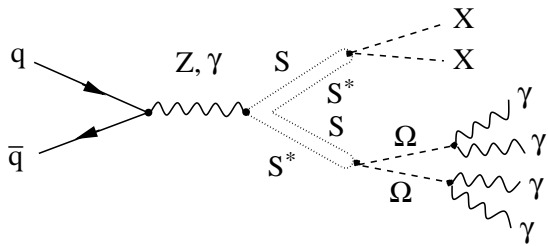}
\caption{Production and decay of $S$-mesons at hadron collider.
}
\vspace{-0.5cm}
\label{collider}
\end{figure}

{\bf Collider Signatures.}  The $S$-mesons would be  produced at LHC
mainly through intermediate $s$-channel photons and $Z$ bosons in the case
where their color pertains to an exotic SU(N) gauge symmetry, as shown in
fig.\ \ref{collider}.   Because of confinement the initially produced
$S$-$S^*$ pair must hadronize to form the  $S$-meson bound states $\phi_S$
($SS^*$) or $\pi_S$ ($S T^*$).  If $m_T\lesssim m_S$ (but not $\ll m_S$),
then $\phi_S$ and $\pi_S$ pairs will form with roughly equal probability. 
In the case of $\pi_S$ production, if the $S$ lifetime is short enough so
that it decays within the detector,  there will be a distinctive signal of
pairs of charged leptons and antileptons, where each like-sign pair has an
invariant mass equal to that of  the $\pi_S$.  

If on the other hand a $\phi_S$ pair is produced, the typical decay
products will be dark matter pairs, $XX$, or glueballs,
$\Omega\,\Omega$, as shown in fig.\ \ref{collider}.  Each glueball
decays into two photons because of the one-loop Euler-Heisenberg 
interaction.  Thus pairs of photons with invariant
mass $m_\Omega$ will be produced, and pairs of pairs will reconstruct to
invariant mass $m_{\phi_S}\cong 2m_S$.  In the example shown, this will be
accompanied by missing energy $m_{\phi_S}$ A detailed study
should be done to see if existing searches of this nature 
\cite{Aad:2011zj} already exclude our model. Another likely
signature would be four pairs of photons due to decay of both mesons
into glueballs. 

{\bf Conclusions.}  We have shown that scalar dark matter $X$ with mass
130 GeV could produce a gamma ray spectral feature tentatively identified
in Fermi-LAT data, with the addition of just one scalar multiplet  $S$
transforming as ($Y_S$,\,1,\,3) under U(1)$_Y\times$SU(2)$_L\times$SU(3),
where the SU(3) might be the gauge group of QCD, or else some new hidden
sector interaction.  The coupling $\lambda_X X^2 |S|^2$ between $S$ and
$X$ need not be very  large unless $q_S \sim 1$ and  $m_S/m_X$ is greater
than a few.  The strong interactions of $S$ serve two purposes: they
confine the stable charged relic component of $S$, and the number of
colors helps to increase the $XX\to\gamma\gamma$ cross section while
keeping $\lambda_X$ reasonably small.

Because $S$ has similar quantum numbers to right-handed squarks, it
is tempting to make this identification, but such light squarks are
ruled out by LHC with the possible exception of the third generation.
Squarks have a smaller electric charge than $S$ in our preferred
examples, and all three generations would need to contribute to
compensate for the resulting decrease in the $XX\to\gamma\gamma$ cross
section.   

It is interesting that we rely upon the $h^2 X^2$ coupling between $X$ and
the Higgs boson to get the thermal relic density of dark matter, and that
the same coupling leads to a cross section for $X$ scattering on nucleons
that is just a factor of 3 below the 2012 Xenon100 direct detection limit. The
model has additional links to Higgs physics: the possibility of increasing
the $h\to\gamma\gamma$ branching ratio by a factor of  a few, and the
existence of bound states of  $S$ and $S^*$ ($\phi_S$), which could have a
large branching ratio to decay into Higgs bosons, though more generically
their decay products are dominated by  glueballs $\Omega$  of the exotic
SU(N) or $X$ bosons.  We suggest that LHC might discover the $S$-mesons,
whose mass should be $>260$ GeV, by observation of two photon pairs each
with invariant mass of $m_{\Omega}$,  accompanied by the same amount of
missing energy, or four photon pairs each of mass $m_{\Omega}$.  In
addition, lighter charged  $S T^*$ mesons ($\pi_S$) should be produced,
decaying into like-sign lepton pairs, which might or might not occur
within the detector. A more detailed study of LHC signals is contemplated
\cite{ip}.

After releasing version 1 of this work, we were informed of a similar
model in \cite{Ibarra:2012dw}, resembling ours in the case where $m_X
\cong m_{\phi_S}$, using the process $XX\to\phi_S\phi_S$ followed by
$\phi_S\to\gamma\gamma$, which becomes the dominant decay channel for
$\phi_S$ if $\lambda_h\ll 1$ and no hadronic channels are available
(as in the case of a hidden SU(N) with glueball mass greater than
$m_{\phi_S}/2$).  This seems to be another viable region of parameter
space for our model, relying only upon tree-level amplitudes. In this
case the DM mass would be 260 rather than 130 GeV.  We acknowledge A.\
Ibarra for pointing this out.

{\bf Acknowledgements}.  I thank C.\ Burgess, A.\ Hektor, A.\ Ibarra, K.\
Kainulainen, Z.\ Liu, M.\ Pospelov, M.\ Raidal, P.\ Scott, M.\ Trott, B.\
Vachon, C.\ Weniger for helpful correspondence or discussions,  and L.\
Vecchi for pointing out an important error concerning electroweak
precision constraints in the first version.  I thank G.\ Moore especially
for many useful discussions about the strong interaction aspects and for
noticing the oversight of baryonic states of $S$ in  the original version
of this work, hence the need for $S$ to be unstable.

JC's research is supported by the Natural Sciences and Engineering
Research Council (NSERC, Canada).

\bibliographystyle{apsrev}

\end{document}